\documentclass[
reprint,
 amsmath,amssymb,
 aps,
]{revtex4-1}

\usepackage{graphicx}
\usepackage{dcolumn}
\usepackage{bm}

\begin{document}

\preprint{PRL}

\title{Preventing Buckling of Slender Cylindrical Structures by Internal Viscous Flows}

\author{Max Linshits}%
\affiliation{Faculty of Mechanical Engineering, Technion - Israel Institute of Technology,
Technion City, Haifa, Israel 32000}
\author{Amir D. Gat}%
\email{amirgat@tx.technion.ac.il}
\affiliation{Faculty of Mechanical Engineering, Technion - Israel Institute of Technology,
Technion City, Haifa, Israel 32000}
\date{\today}

\begin{abstract}
Viscous flows within an elastic structure apply stress on the solid-liquid interface. The stress-field created by the viscous flow can be utilized to counter stress created by external forces and thus may be applied as a tool for delaying the onset of structural failure. To illustrate this concept we study viscous flow within an elastic cylinder under compressive axial force. We obtain a closed-form expression showing an approximately linear relation between the critical buckling load and the liquid inlet pressure. Our results are validated by numerical computations.
We discuss future research directions of fluid-solid composite materials which create flow under external stress, yielding enhanced resistance to structural failure.

\end{abstract}

\pacs{Valid PACS appear here}
\maketitle

Viscous flows within an elastic solid apply both pressure and shear stress at the solid-liquid interface and thus create an internal stress- and deformation-fields within the solid. The stress-field created by the viscous flow can be utilized to counter stress-field created by external forces, similarly to pre-stressing of solid structures \cite{Raju.2006}, and thus may be applied as a tool for delaying the onset of structural failure. We illustrate this concept by examining the effect of internal viscous flow on buckling of an elastic cylinder under compressive axial force. The examined case is relevant to applications such as soft-robots and soft-actuators \citep{elbaz2014dynamics,trivedi2008soft} and micro-needles \cite{Park.2005, Sullivan.2010,Hood.2011}, where the buckling failure mode is one of the limiting factors determining the needle radius and thus is of interest for on-going efforts to create less invasive, painless micro-needle arrays.

Various existing works studied the interaction between external viscous flow and elastic buckling of a slender filaments. Becker \& Shelley (2001) \cite{becker2001instability} studied deformation of a slender elastic filament due to a flow-field with uniform shear stress and examined bifurcations of the deformation shape. Wandersman \textit{et al.} (2010) \cite{wandersman2010buckled} conducted experiments of elastic fibers within a viscous flow composed of an array of vortices. Guglielmini \textit{et al.} (2012) \cite{guglielmini2012buckling} studied, analytically and numerically, the effect of external viscous flow on the buckling instability of a filament clamped to a rigid wall. Buckling and instabilities of elastic cylinders due to internal viscous flow was extensively studied in the context of pipes conveying fluid and collapsible tubes \cite{Heil.1996a,Heil.1996b,paidoussis1998fluid}.

\begin{figure}[b]
\includegraphics[width=0.45\textwidth]{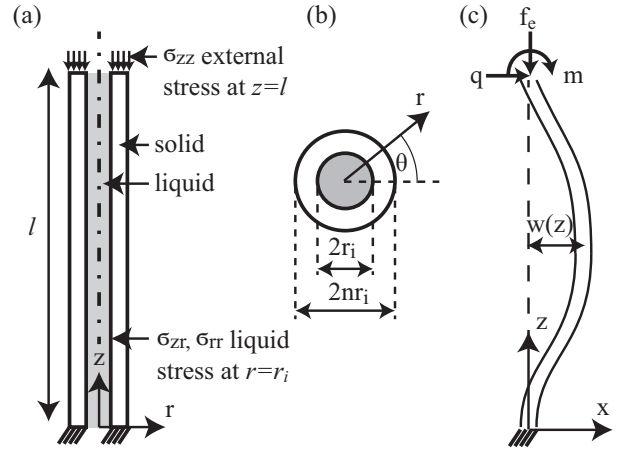}
\caption{A schematic description of the stress applied on the cylinder in the unbuckled state (a) and the cylinder cross section (b). Part (c) presents the deflection and forces and moment acting at $z=l$ for the buckled case.}
\label{figure_1}
\end{figure}

In this work we focus on steady viscous, Newtonian, incompressible flow through a slender linearly elastic cylinder (see Figure \ref{figure_1}) and study the effects of shear stress applied by the liquid on the onset of the buckling failure mode due to external axially compressible force. Hereafter, capital letters denote normalized variables and asterisk superscript denote characteristic values. We denote the length of the cylinder $l$, the inner radius $r_i$, the outer radius $r_o$, the Young's modulus of the solid $e_m$, the critical buckling load $f_c$, the second moment of inertia of the solid cross section $m_i$, the liquid viscosity $\mu$, liquid density $\rho$, liquid speed in the axial direction $w$, characteristic solid radial deformation $d_z^*$, characteristic solid axial deformation $d_z^*$ and characteristic solid stress $\sigma_{ij}^*$ (acting on the plane normal to coordinate $i$ and in the direction of coordinate $j$). 

We require small radius-to-length ratio 
\begin{equation}
\frac{r_i}{l}=\varepsilon_1 \ll 1
\end{equation}
and small reduced Reynolds number
\begin{equation}
Re_r=\varepsilon_1 \frac{\rho w^* r_i}{\mu} \ll 1,
\label{Rey}
\end{equation}
which is the relevant parameter expressing the ratio of inertia to viscosity in slender geometries, where $w^*$ is the characteristic velocity. We estimate the characteristic pressure in the liquid by comparing the order-of-magnitude of the critical buckling load $f_e$ and the total axial force applied the by the liquid on the solid
\begin{equation}
f_e= \frac{e_m m_i}{l^2}\sim p^* \pi r_i^2 
\end{equation}
where $k$ is effective column length factor and thus $p^*=e_m m_i/l^2 \pi r_i^2$. 

We define the normalized coordinates
\begin{equation}
Z=\frac{z}{l},\,\,\,R=\frac{r}{r_i},\,\,\, F_e=\frac{f_e  l^2}{ e_m m_i},\,\,\,P=\frac{p l^2 \pi r_i^2}{e_m m_i},
\end{equation}
corresponding to the normalized axial coordinate, normalized radial coordinate, external force and liquid pressure, respectively. Under the above assumptions the flow within the cylinder is Hagen-Poiseuille flow and thus the stress applied by the liquid on the solid surface $r=r_i$ is
\begin{equation}
	\Sigma_{zr}(Z,R=1)=\frac{\Delta P}{2},\,\,\,\Sigma_{rr}(Z,R=1)=P(0)+Z\Delta P,
\end{equation}
where $\Delta P=P(1)-P(0)$ is the normalized pressure difference, $\Sigma_{zr}$ is the dimensionless shear stress applied by the liquid, normalized by $\sigma^*_{zr}=\varepsilon_1 p^*$ and $\Sigma_{rr}$ is the dimensionless radial stress, normalized by $\sigma^*_{rr}=p^*$. Substituting the Hagen-Poiseuille relation $w^*=r_i^2 p^*/8\mu l$ (where $w^*$ is estimated as the average liquid speed) and $m_i=\pi r_i^4 (n^4-1)/4$ into (\ref{Rey}) yields
\begin{equation}
Re_r=\varepsilon_1 \frac{\rho  r_i^2 p^* r_i}{8 \mu^2 l}=\varepsilon_1^4 \frac{\rho  e_m  r_i^2 (n^4-1)}{32 \mu^2} \ll 1.
\end{equation}

We can thus estimate the characteristic stresses $\sigma_{rr}^*$, $\sigma_{zz}^*$, $\sigma_{\theta\theta}^*$ within the cylinder by 
\begin{equation}
\sigma_{rr}^*=\frac{ e_m m_i\varepsilon_1}{l^2 \pi r_i^2},\quad  \sigma_{zz}^*=\frac{ e_m m_i}{l^2 \pi r_i^2 (n^2-1)},
\end{equation}
and
\begin{equation}
\sigma_{\theta\theta}^*=\sigma_{rr}^* \frac{1}{n-1}=\frac{e_m m_i\varepsilon_1}{l^2 \pi r_i^2 (n-1)}.
\end{equation}
From the constitutive relation and Hook's law we obtain
\begin{equation}
\frac{\partial d_z}{\partial z}=\frac{\sigma_{zz}-\nu\left(\sigma_{rr}+\sigma_{\theta\theta}\right)}{e_m},\,\,\,
\frac{d_r}{r}=\frac{\sigma_{\theta\theta}-\nu\left(\sigma_{rr}+\sigma_{zz}\right)}{e_m}
\end{equation}
and since $\sigma_{zz}^*\gg\sigma_{zr}^*,\sigma_{rr}^*,\sigma_{\theta\theta}^*$ we can estimate the deformation field based on $\sigma_{zz}^*$ as
\begin{equation}
\frac{d_z^*}{l}\sim\frac{\varepsilon_1^2}{4}(n^2+1) \ll 1,\,\,\,\frac{d_r^*}{r_i}\sim \nu \frac{d_z^*}{l} \ll 1
\end{equation}
and thus the cylinder deformations are negligible for the forces and pressures characteristic with buckling. We can thus limit our analysis to cases with constant channel cross-section. 

We follow the classic approach by Euler (1757) \cite{Euler.1757,Paidoussis.1998} and calculate the minimal value of the external force $F_e$ for which a solution of the bending equation of the beam can be obtained. The governing equation is
\begin{equation}
e_m m_i \frac{\partial^2 w}{\partial z^2}+\left[f_e-\Delta p \pi r_i^2 \left(1-\frac{z}{l}\right)\right]w=q(l-z)+m.
\end{equation} 
We define the normalized parameters
\begin{equation}
W=\frac{w}{w^*},\,\,\, Q=\frac{q k^2 l^3}{e_m m_i w^*},\,\,\,M=\frac{m k^2 l^3}{e_m m_i w^*}
\end{equation}
representing, deflection, lateral reaction force at $Z=1$ and moment at $Z=1$, respectively. The normalized equation is thus
\begin{equation}
\frac{\partial^2 W}{\partial Z^2}+[F_e -\Delta P(1-Z)]W=Q(1-Z)-M.
\end{equation}
Substituting $\zeta=\Delta P^{-2/3}(F_e-\Delta P(1-Z))=\Delta P^{-2/3}F_e-\Delta P^{1/3}(1-Z))$ we obtain the Airy equation and the homogenous solution $Y_H$ is thus
\begin{equation}
Y_H=C_1 A(\zeta)+C_2 B(\zeta)
\end{equation}
where $A$ and $B$ are the zero and second Airy functions, respectively. The complete solution can be obtained by variation of parameters as (utilizing the relation $A(\zeta)B'(\zeta)-A'(\zeta)B(\zeta)=1/\pi$)   
\begin{eqnarray}
W(Z)=\pi B\frac{F_e-\Delta P(1-Z)}{\Delta P^{2/3}} \bigg[C_1+\\
\int_0^Z{(M+Q-Q\xi)A\left(\frac{F_e-\Delta P(1-\xi)}{\Delta P^{2/3}}\right)d\xi}\bigg] \nonumber \\
-\pi A\frac{F_e-\Delta P(1-Z)}{\Delta P^{2/3}} \bigg[C_2+\nonumber \\
\int_0^Z{(M+Q-Q\xi)B\left(\frac{F_e-\Delta P(1-\xi)}{\Delta P^{2/3}}\right)d\xi}\bigg]. \nonumber
\end{eqnarray}
The boundary conditions at $Z=0$ are
\begin{equation}
W(Z=0)=0,\,\,\,\frac{\partial W(Z=0)}{\partial Z}=0,
\label{Z_zero_BC}
\end{equation} 
yielding $C_1=C_2=0$. The boundary conditions at $Z=1$ can thus be described as functions of $Q$ and $M$
\begin{eqnarray}
W(1)=\pi \int_0^1{\left(M+Q-Q\xi\right)}\times \\
\bigg[B\left(\frac{F_e}{\Delta P^{2/3}}\right)A\left(\frac{F_e-\Delta P(1-\xi)}{\Delta P^{2/3}}\right)\nonumber \\
- A\left(\frac{F_e}{\Delta P^{2/3}}\right)B\left(\frac{F_e-\Delta P(1-\xi)}{\Delta P^{2/3}}\right)\bigg]d\xi \nonumber
\label{W1}
\end{eqnarray}
and
\begin{eqnarray}
\frac{\partial W(1)}{\partial Z}=\pi \Delta P^{1/3} \int_0^1{\left(M+Q-Q\xi\right)}\times \\
\bigg[B'\left(\frac{F_e}{\Delta P^{2/3}}\right)A\left(\frac{F_e-\Delta P(1-\xi)}{\Delta P^{2/3}}\right)\nonumber \\
- A'\left(\frac{F_e}{\Delta P^{2/3}}\right)B\left(\frac{F_e-\Delta P(1-\xi)}{\Delta P^{2/3}}\right)\bigg]d\xi. \nonumber
\label{dW1}
\end{eqnarray}

\begin{figure}
\includegraphics[width=0.5\textwidth]{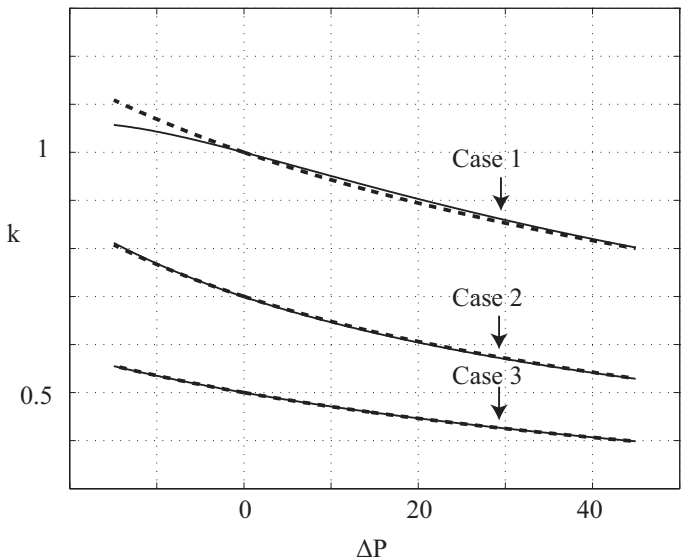}
\caption{\label{fig:illustration} The value of $k$, effective column length factor, as function of $\Delta P$. For all cases the column is fixed at $Z=0$. At $Z=1$ $M=0$ and $W(1)=0$ for case 1,  $M=0$ and $W(1)=0$ for case 2 and $\partial W (1)/\partial Z=0$ and $W(1)=0$ for case 3. The smooth lines are solutions of (\ref{W1})-(\ref{dW1}) and the dashed lines are a approximated solution (\ref{ApproxSol}).}
\end{figure}

These relations enable solving for the minimal value of $F_e$ for various boundary conditions and liquid pressure differences $\Delta P$. For the case of fixed end at $Z=1$, $\partial W (1)/\partial Z=0$ and $W(1)=0$, (\ref{W1}) and (\ref{dW1}) are solved for $F_e$ and $Q/M$ simultaneously. For hinged end at $Z=1$, $M=0$ and $W(1)=0$, (\ref{W1}) is solved for $F_e$. For end free to move laterally at $Z=1$, $Q=0$ and $\partial W (1)/\partial Z=0$, (\ref{dW1}) is solved for $F_e$.

We approximate the obtained solutions to the linear expression
\begin{equation}
F_e \approx \frac{\pi^2}{k^2_0} \left(1+C\Delta P\right)
\label{sol}
\end{equation}
or in dimensional form as $f_c=\pi^2/k_0^2(e_m m_i /l^2 +C\pi r_i^2\Delta p)$, where $k_0$ is the effective column length factor calculated for the case without internal flow and $C$ is a coefficient representing the effect of the viscous flow. The value of $C$ is $C=1/80$ for $\partial W (1)/\partial Z=0$ and $W(1)=0$, $C=1/60$ for $\partial W (1)/\partial Z=0$ and $M=0$, and $C=1/80$ for $\partial W (1)/\partial Z=0$ and $Q=0$. Equivalently, we can present this approximation in terms of effective column length factor $k$ as
\begin{equation}
k \approx \frac{k_0}{\sqrt{1+C \Delta P}}=\frac{k_0}{\sqrt{1+C \pi r_i^2\Delta p l^2/e_m m_i}}.
\label{ApproxSol}
\end{equation}
In figure \ref{fig:illustration} we present the value of $k$, effective column length factor, as function of $\Delta P$ for various configurations. For all cases the column is fixed at $Z=0$. At $Z=1$ $M=0$ and $W(1)=0$ for case 1,  $M=0$ and $W(1)=0$ for case 2 and $\partial W (1)/\partial Z=0$ and $W(1)=0$ for case 3. The smooth lines are solutions of (\ref{W1})-(\ref{dW1}) and the dashed lines are a approximated solution (\ref{ApproxSol}). Good agreement between the full and approximated solution is evident and a significant monotonic decrease in $k$, and thus an increase in rigidity, as $\Delta P$ increases.

\begin{figure}
\includegraphics[width=0.45\textwidth]{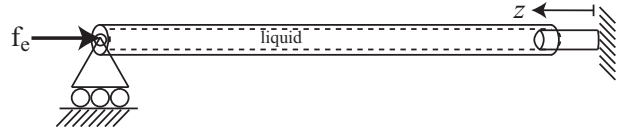}
\caption{An illustration of a simple solid-liquid composite material. A small pin with length small compared
with the length of the cylinder is connected to the base of a liquid-filled cylinder and thus flow is created
when external force is applied.}
\label{figure_3}
\end{figure}



Our results can be readily applied as a tool to temporarily increase the buckling rigidity of cylindrical structures. Future research can build on our results to design a composite solid-liquid material which combines an external elastic structure and channels containing viscous fluid. Deformation of such a structure under external force can create viscous flows which apply stress on the solid and thus significantly change the stress-field in the structure which may allow
delaying structural failure and gaining complex structural deformations. An illustration of a simple solid-liquid composite material is presented in figure \ref{figure_3}. A small pin with length small compared with the length of the cylinder is connected to the base of a liquid-filled cylinder and thus flow is created when external force is applied. The cylinder length is $l$, outer radius $r_o$, inner radius $r_o/2$, and Young's modulus $e_m$. The critical buckling load of such configuration (see (\ref{sol})) is $24.95 (e_m \pi r^4)/l^2$  while the critical buckling load of a filled solid cylinder is $20.19 (E \pi r^4)/l^2$. Hence, by removing solid and replacing it with liquid an increase to the buckling load was obtained.


%

\end{document}